\documentclass[10pt,letterpaper,twocolumn]{article} %% two column, final layout

\usepackage{ol2}
\usepackage[draft]{hyperref}
\usepackage{amsmath}
\usepackage{amssymb}

\begin{document}

\twocolumn[%% activate for two-column option

\title{Programming scale-free optics in disordered ferroelectrics}

%% For REVTeX it is possible to automate superscript and e-mail callouts with the superscriptaddress option; see REVTeX4 documentation.

\author{Jacopo Parravicini,$^{1,2}$ Claudio Conti,$^{1,3}$ Aharon J. Agranat, $^{4}$ Eugenio DelRe $^{1,2}$}

\address{$^1$ Department of Physics, Universit\`{a} di Roma ``La Sapienza'',
00185 Roma, Italy \\
$^2$IPCF-CNR, Universit\`{a} di Roma ``La Sapienza'', 00185 Roma, Italy
\\
$^3$ISC-CNR, Universit\`{a} di Roma ``La Sapienza'', 00185 Roma, Italy
\\
$^4$Applied Physics Department, Hebrew University of Jerusalem,
91904 Israel
}

\begin{abstract}
Using the history-dependence of a dipolar glass hosted in a compositionally-disordered lithium-enriched potassium-tantalate-niobate (KTN:Li) crystal, we demonstrate scale-free optical propagation at tunable temperatures. The operating equilibration temperature is determined by previous crystal spiralling in the temperature/cooling-rate phase-space.
\end{abstract}

\ocis{160.2750, 190.4400, 190.5330}

] %% activate for two-column option

Optical diffraction can be compensated in waveguides or when solitons form. The waves must have a specific size, shape, and specific intensity, for the solitons. Recent experiments in electromagnetically-induced-transparency and nonlinear optics have identified a situation in which diffraction is compensated, as for solitons, but without the scale-dependent constraints, a phenomenon termed \emph{diffraction cancellation} \cite{Firstenberg2009,DelRe2011}. In photorefractive crystals this has been demonstrated in a regime of so-called ``scale-free optics'' supported by the formation of a dipolar glass, a phenomenon typical of disordered ferroelectric cystals that host glass-forming polar nanoregions (PNRs) when rapidly cooled\cite{Samara2003,Bokov2006}. Scale-free optics opens the way to a number of enticing effects, such as wavelength-insensitive propagation \cite{Parravicini2011} and scale-free spatial instability \cite{Folli2012}. Moreover, for its range of validity \cite{Note2}, the underlying model  predicts a regime that supports subwavelength beam propagation \cite{Conti2011}.

Experimentally, rapid cooling gives rise to scale-free optics only when the cooling rate is above a threshold value and the final equilibration temperature coincides with the Curie point $T_C$ (in cooling), where the paralectric-ferroelectric phase transition occurs \cite{DelRe2011}. In fact, the scale-free model requires the photorefractive diffusion field to cancel diffraction, a phenomenon that entails an enhanced static dielectric response as only observed at $T_C$ \cite{Jona1993}. The key is that the dielectric anomaly is rendered accessible by the PNR-driven glass which suppresses long range order and the optical scattering typical of the equilibrium phase-transition \cite{Kleemann1997,DelRe1999}. %

\begin{figure}[t!]
\begin{center}
\includegraphics[width=0.75\columnwidth]{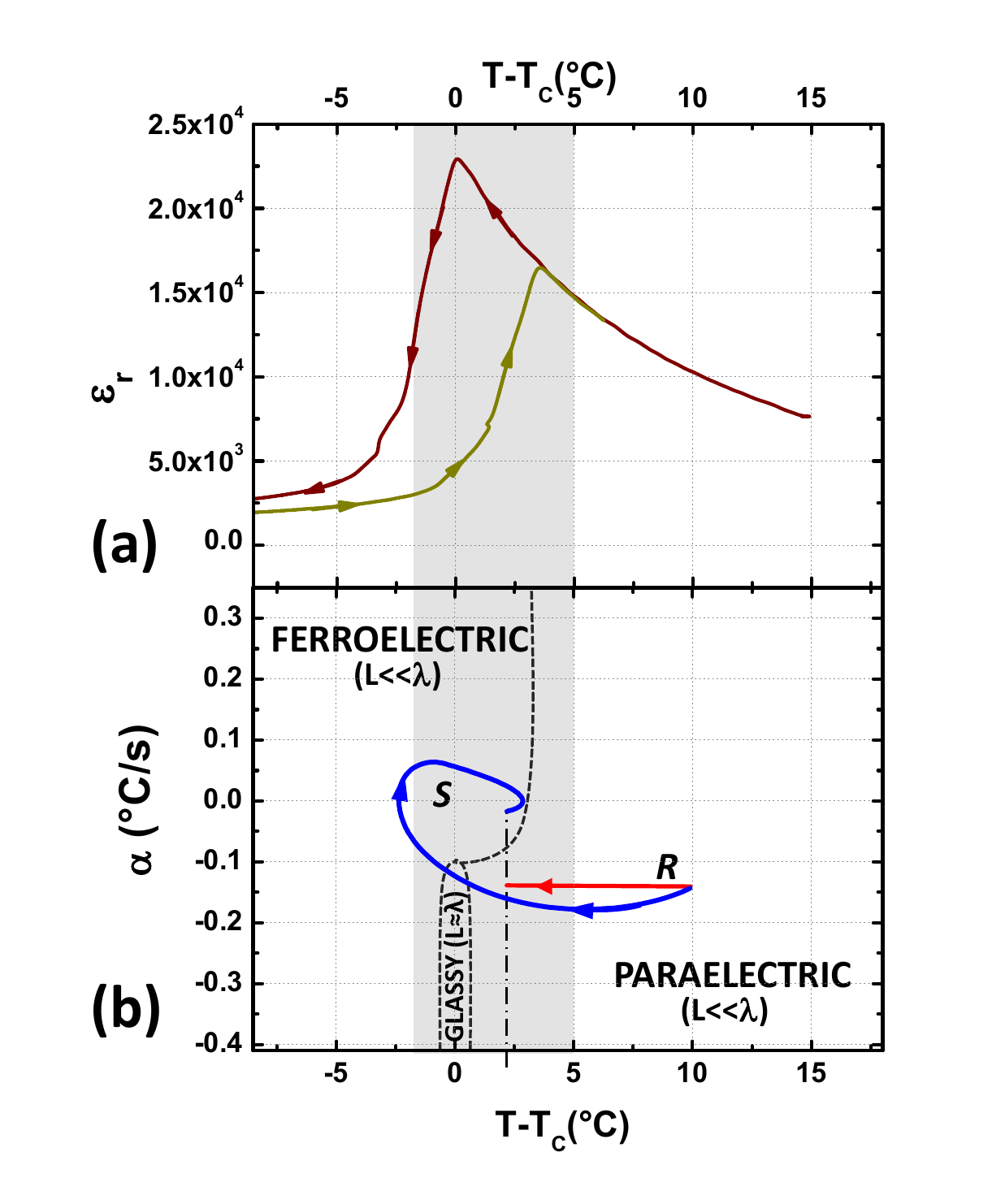}
\caption{(a) KTN:Li static dielectric constant $\varepsilon_r$  for slow cooling (red curve) and slow heating (green curve) as a function of $T$. (b) Representation of the parameter plane $(T-T_C,\alpha)$ of the sample preparation. Both regions of paraelectric and ferroelectric behavior, which are unable to support scale-free propagation through rapid cooling, are indicated, together with the glassy region (dashed line indicates the approximate phase separation). The shaded region indicates the range of hysteresis, while the ``\textsc{Glassy}'' region indicates the temperature-cooling rate range where standard constant $\alpha$ preparation (rapid cooling) can give rise to to $L\simeq\lambda$ (see text). The red (R) and blue (S) curves represent respectively the rapid-cooling ($\alpha=\text{constant}$) and the spiralling trajectories with initial ($T_i = 24.5$ $^\circ$C) and final ($T_f = 16.8$ $^\circ$C, dot-dashed line) temperatures that refer to the experimental data in Fig. \ref{Figure2}.}
\label{Figure1}
\end{center}
\end{figure}

In PNR-driven dipolar-glasses thermal history can actually shift the crystal $T_C$ with its associated anomalous enhancement of response. This property gives rise to a marked thermal hysteresis that delimits  what is termed the crystal glassy phase (see Fig. \ref{Figure1}a) \cite{Bokov2006,Leuzzi2008,Mossa2004,Parravicini2012}.  In this letter we program the temperature at which scale-free optics occurs by exploiting specific sequences of heating and quenching stages.  In our results, we are able to span the greater part of the crystal glassy phase.

To grasp the physical underpinnings of our experiments, we recall that photorefraction leads to a diffusive nonlinearity \cite{Crosignani1998,Crosignani1999}, which profoundly alters beam propagation, in that diffraction is governed by an effective refractive index $n_\text{eff}$ \cite{DelRe2011, Conti2011,Parravicini2011}
\begin{equation}\label{neff}
n_\text{eff}=n_0/(1-\left(L/\lambda\right)^2),
\end{equation}
where $n_0$ is the unperturbed refractive index, $\lambda$ the wavelength and $L=4 \pi n_0^2  \varepsilon_0 \sqrt{g} \chi_{\text{PNR}} (K_B T/q)$. Here $g$ is the effective quadratic electro-optic coefficient, $\chi_\text{PNR}$ is the effective history-dependent low-frequency dielectric susceptibility of the dipolar glass, $K_B$ the Boltzmann constant, $T$ the crystal equilibration temperature (i.e. the temperature measured at a given istant) and $q$ is the charge of the photoexcited carriers. Eq. (\ref{neff}) is valid for $L\lesssim\lambda$. As $L \rightarrow \lambda$, $n_\text{eff} \gg n_0$ and diffraction is cancelled, the scale-free regime, independently of beam size and intensity. The condition $L=\lambda$ requires $\chi_\text{PNR}\sim 10^5$ (which imposes $T=T_C$ for rapid cooling schemes).

We use a 6$\times$3$\times$2.5mm sample of  ferroelectric Li-enriched Cu-doped potassium-tantalate niobate (KTN:Li). In Fig. \ref{Figure1}a we report the quasi-static dielectric response $\varepsilon_r$ versus $T-T_C$ ($T_C=14.5 ^{\circ}$C) obtained from capacitance measurements at slow heating and cooling rates ($|\alpha|=|\Delta T/\Delta t| \simeq 0.01$ $^\circ$C/s) \cite{Parravicini2012}. The marked hysteresis in $\varepsilon_r(T-T_C)$ (shaded region in Fig. \ref{Figure1}) signals a non-ergodic phase several degrees above and below $T_C$, a well-known complex dielectric response \cite{Ishai2004,Gumennik2011} that is analogous to that observed in other relaxors \cite{Bokov2006}. The Cu doping gives it a greenish tint and a strong photorefraction \cite{DelRe2011}.

Thermal preparation required to morph $\chi_\text{PNR}$ is achieved through a computer-controlled Peltier-junction that fixes the temperature versus time $T=T(t)$ schedule (the sequence of quenching and heating preparatory stages). Once the thermal preparation has been completed, we shine laser light into the sample and detect diffraction. We focus a  $\lambda=633 \ \mathrm{nm}$ TEM$_{00}$ linearly polarized (in the horizontal $x$-direction) $z$-propagating beam onto the input facet and analyze the input and output beam intensity distribution Full-Width-at-Half-Maximum (FWHM) using an imaging system and a CCD camera. The condition $L = \lambda$ is found when input and output FWHM measurements coincide, i.e., the diffraction of micron-sized beams ceases, independent of size and peak intensity.

\begin{figure}
\begin{center}
\includegraphics[width=0.9\columnwidth]{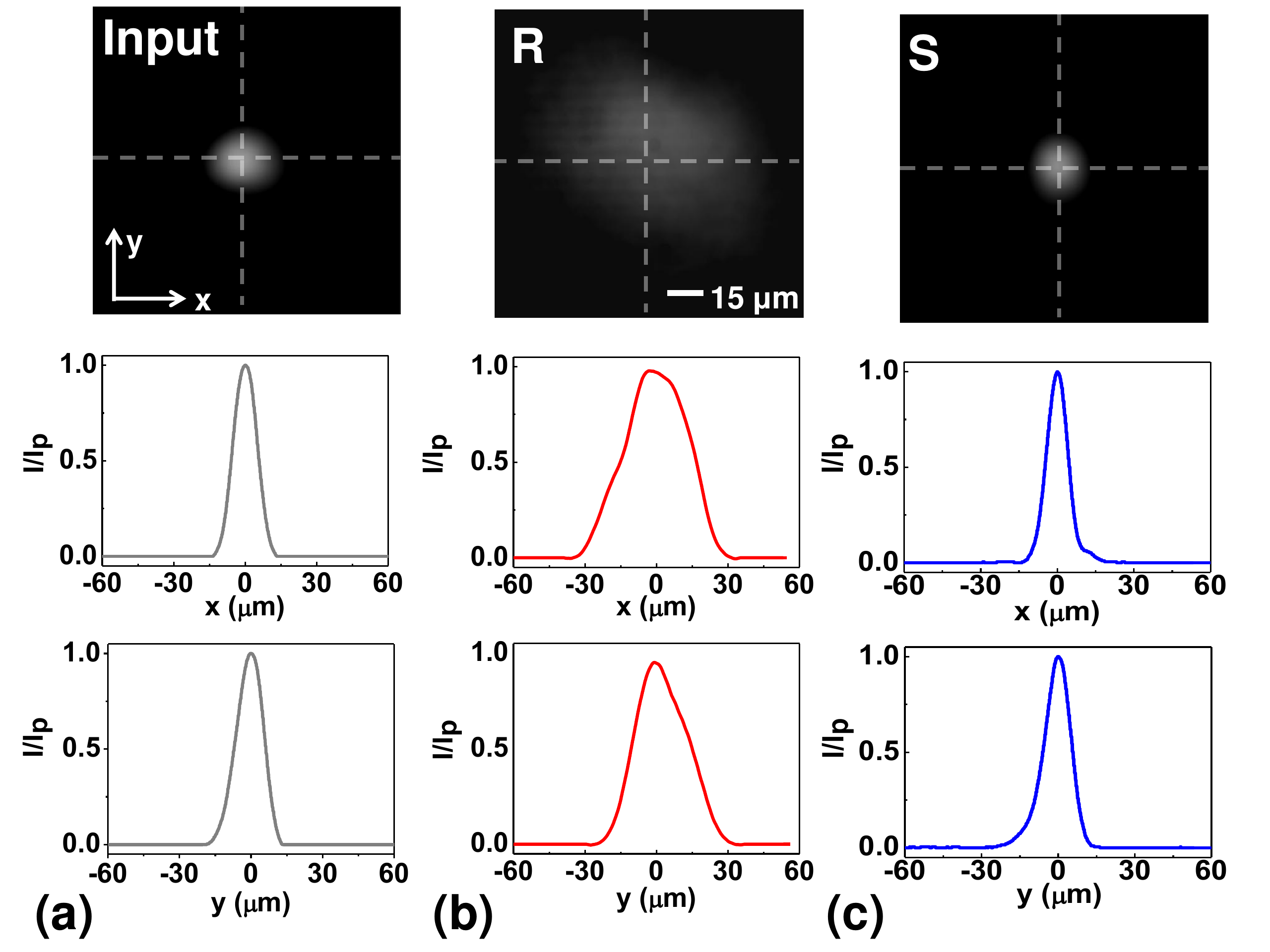}
\caption{Intensity distributions and $x$ and $y$ profiles of the beam in the conditions of rapid cooling (R) and spiralling (S) of Fig.\ref{Figure1}b. (a) Input facet of the sample ($\Delta x \simeq \Delta y \simeq 12 \ \mathrm{\mu m}$); (b) output facet after the rapid-cooling process  (curve $R$, $L/\lambda\ll 1$); (c) output facet after the spiralling thermal trajectory (curve $S$, $L=\lambda$). Note that (b) manifests a quasi-linear diffraction ($\Delta x \simeq \Delta y \simeq 33 \ \mathrm{\mu m}$), while the intensity profile in (c) is devoid of spreading ($\Delta x \simeq \Delta y \simeq 12 \ \mathrm{\mu m}$).}%
\label{Figure2}
\end{center}
\end{figure}

We first carry out experiments using the standard constant $\alpha$ thermal preparation (rapid cooling). As reported previously \cite{DelRe2011}, only a small region in the experimentally available parameter space $(T-T_C,\alpha)=(T-T_C,\dot{T})$ gives rise to $L\simeq\lambda$, tagged in Fig. \ref{Figure1}b as``\textsc{Glassy}''. The region is evidently smaller compared to the thermal range in which dielectric hysteresis is detected (shaded region). A typical result for parameters external to the glassy region is shown in Fig. \ref{Figure2}b. Here, the rapid cooling process is characterized by $(T-T_C,\alpha)=(2.3^{\circ}C,-0.12^{\circ}C/s)$. As seen in the transverse intensity distribution images, the input intensity distribution with a FWHM of the $\Delta x \simeq \Delta y \simeq 12 \ \mathrm{\mu m}$ (see Fig. \ref{Figure2}a) diffracts and spreads to $\Delta x \simeq \Delta y \simeq 30 \ \mathrm{\mu m}$ after the $l_z =3$ mm propagation in the sample (Fig. \ref{Figure2}b). As expected, diffraction cancellation is not observed. Comparing  diffraction to expected Gaussian beam diffraction ($\Delta x_\text{out}\sim 33$ $\mu$m, given the crystal $n_0$=2.38 and $\lambda = 633$ nm), we find $L/\lambda\lesssim 0.1$ ($n_\text{eff}\simeq n_0$). We consider paraxial diffraction where a standard launch/detection scheme is sufficient (see \cite{DelRe2011}).

We next repeat the experiment reaching the same final temperature ($T-T_C = 2.3$ $^\circ$C), but through a different thermal preparation, characterized by a non-monotonic thermal path. The set of successive points occupied by the sample in the $(T-T_C,\alpha)$ plane at the different instants of time is the curve $S$ in Fig. \ref{Figure1}b. This representation allows to appreciate how, in terms of the rapid cooling experiments, the crystal is actually exposed in sequence to all three ``phases'', the paraelectric, the glassy, and the ferroelectric one. After this preparation, the output beam intensity distribution, reported in Fig. \ref{Figure2}c, manifests no diffraction, signalling $L/\lambda\simeq 1$ even though the final equilibration temperature is external to original \textsc{Glassy} region (but still contained in the shaded one). We thus conclude that the spiralling trajectory S in the ($T-T_C,\alpha$) plane is able to produce the giant $\chi_\text{PNR}$ which is necessary to obtain $L/\lambda\simeq 1$ even though the final temperature $T\neq T_C$. For comparative purposes, in Fig. \ref{Figure1}b we also plot the $R$ curve associated to the rapid cooling process (of Fig. \ref{Figure2}b) \cite{Note1}.

\begin{figure}
\begin{center}
\includegraphics[width=0.95\columnwidth]{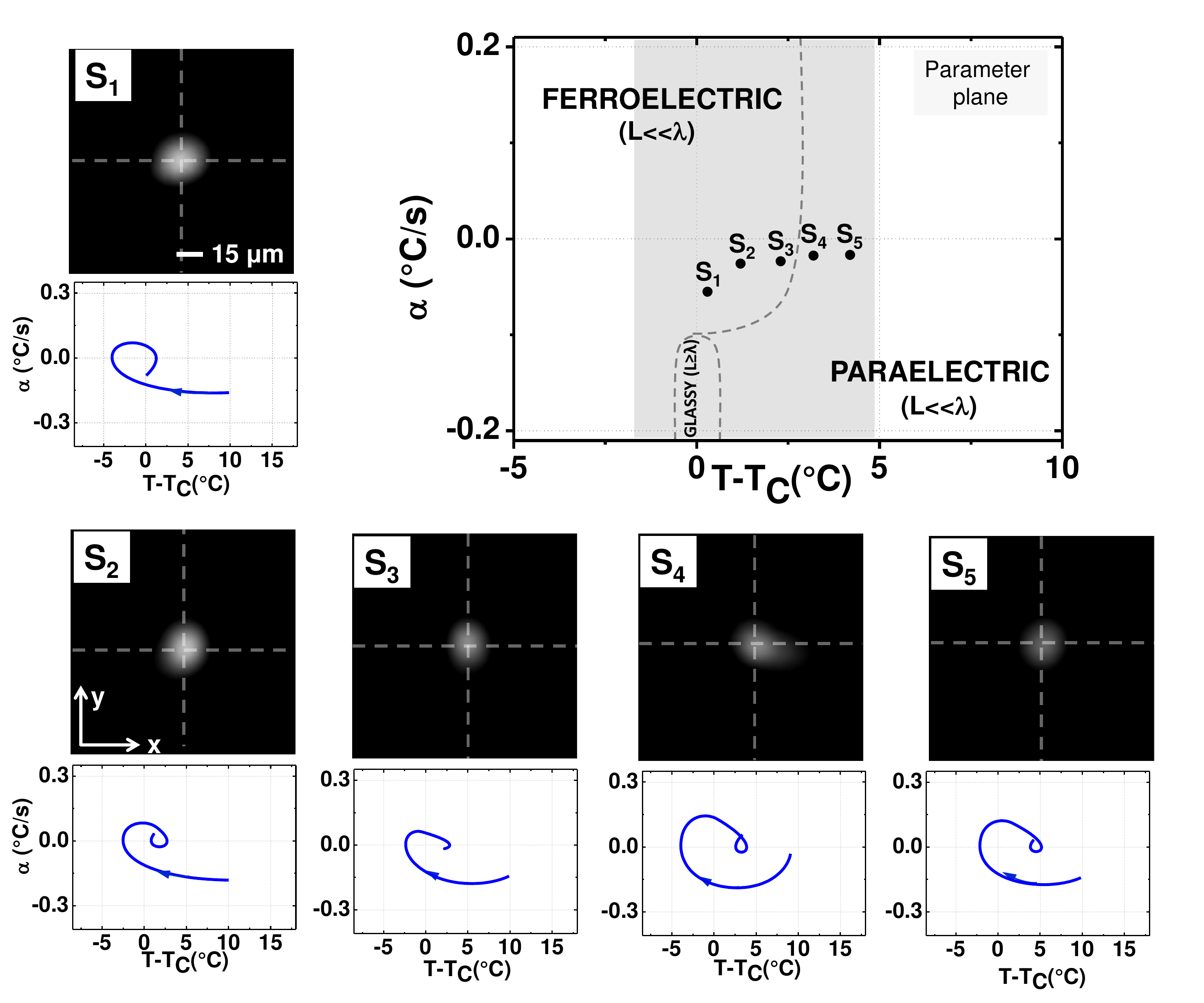}
\caption{Programming scale-free response at temperatures $T\neq T_C$. (Top right) Representation in the $(T-T_C,\alpha)=(T-T_C,\dot{T})$ parameter plane of the different conditions of scale-free optics reported. Activation, testified by the diffraction-free output intensity distribution (top) reported in the series of images S$_{1-5}$, is achieved through spiralling thermal trajectories (bottom). Programmed temperatures $T\simeq$ $14.8$ $^\circ$C (S$_1$), $15.7$ $^\circ$C (S$_2$), $16.8$ $^\circ$C (S$_3$), $17.7$ $^\circ$C (S$_4$), $18.7$ $^\circ$C (S$_5$). In the parameter plane representation, $\alpha$ is the mean value along the trajectory. The points lie in a temperature range of more than $4$ $^\circ$C.}%
\label{Figure3}
\end{center}
\end{figure}

We next extend the use of these ``spiralling'' trajectories to obtain a scale-free response (i.e., excited glassy behavior) throughout the shaded region in Fig. \ref{Figure1}. Results reported in Fig. \ref{Figure3} demonstrate the activation of scale-free propagation regime at tunable temperatures beyond the \textsc{Glassy} region.
We underline that our results are associated to aging and are hence transient in time. For a beam with a peak intensity of $I_{p} \simeq 6 \ \mathrm{W/cm}^2$, the scale-free phenomenon occurs after an exposure of 20 s and begins decaying after approximately 130 s. Remarkably, once the aging has washed out the scale-free regime, the response can be rejuvenated at the same or at a different temperature. Repeating the full cycle, i.e., washing out previous photorefractive charge displacement (resetting of the dipolar glass), the same trajectory will lead to the same optical response.

In terms of soft-matter physics, our experiments demonstrate a method to outdo the strong temperature selectivity required in rapid cooling experiments through the complexity of the dipolar glass in the region of the non-ergodic phase, identified by  thermal hysteresis in the dielectric susceptibility \cite{Samara2003,Bokov2006}. Our technique, based on the use of non-monotonic thermal preparation, appears analogous to the cross-over or Kovacs effects in soft-matter \cite{Leuzzi2008,Mossa2004}. Triggered  by non-monotonic thermal preparation, cross-over is known to lead to effects that are absent through rapid cooling \cite{Parravicini2012}.

In summary, we have demonstrated how to use cross-over and aging effects to tune the specific temperature at which scale-free optics is obtained in a disordered ferroelectric. The result is a first direct example of programmable optics, i.e. of how complexity can be used to achieve a desired response without resorting to material engineering. Loosening temperature selectivity simplifies further experimental endeavors, such as the all-important search for sub-micron beam propagation \cite{Note2,Conti2011}, where spatial solitons intrinsically break down \cite{DelRe2006}.

We thank M. Deen Islam for technical assistance. This work was supported by funding from the Italian Ministry of Research through the \textsc{Firb} grant \textsc{Phocos}-RBFR08E7VA and from the ERC under the European Community 7th Framework Program (FP7/2007-2013)/ERC grant agreement no. 201766. Partial funding was received by the \textsc{SmartConfocal} of the Regione Lazio and by the \textsc{Prin} no. 2009P3K72Z projects. A.J.A. acknowledges the Peter Brojde Center for Innovative Engineering.

%%%%%%%%%%%%%%%%%%%%%%%%%%%%%%%%%%%%%%%%%%%%%%%%

%\clearpage


\begin{thebibliography}{99}

\bibitem{Firstenberg2009} O. Firstenberg, P. London, M. Snuker, A. Ron, N. Davidson, ``Elimination, reversal and directional bias of optical diffraction'', \emph{Nat. Physics} \textbf{5}, 665 (2009).

\bibitem{DelRe2011} E. DelRe, E. Spinozzi, A.J. Agranat, C. Conti, ``Scale-free optics and diffractionless waves in nanodisordered ferroelectrics'' \emph{Nat. Photonics} \textbf{5}, 39 (2011).

\bibitem{Samara2003} G.A. Samara, ``The relaxational properties of compositionally disordered ABO$_3$ perovskites'', \emph{J. Phys. Condens. Matter} \textbf{15}, R367 (2003).

\bibitem{Bokov2006} A. Bokov, ``Recent progress in relaxor ferroelectrics with perovskite structure'', \emph{J. Mater. Sci.} \textbf{41}, 31 (2006).

\bibitem{Parravicini2011} J. Parravicini, F. Di Mei, C. Conti, A.J. Agranat, E. DelRe, ``Diffraction cancellation over multiple wavelengths in photorefractive dipolar glasses'', \emph{Opt. Express} \textbf{19}, 24109 (2011).

\bibitem{Folli2012} V. Folli, E. DelRe, C. Conti, ``Beam Instabilities in the Scale-Free Regime'' \emph{Phys. Rev. Lett.} \textbf{108}, 033901 (2012).

\bibitem{Note2} The model breaks down for scales at which the space charge saturates. This is a sample-dependent limit associated to the acceptor density $N_A$ and the static $\varepsilon_r$. In typical KTN:Li samples $N_A \simeq 2\cdot 10^{18}$ cm$^{-3}$, so that even in the crytical regime with $\varepsilon_r \sim 10^5$, the model breaks down for beam widths $l\sim 0.1$ $\mu$m.

\bibitem{Conti2011} C. Conti, A.J. Agranat, E. DelRe, ``Subwavelength optical spatial solitons and three-dimensional localization in disordered ferroelectrics: Toward metamaterials of nonlinear origin'', \emph{Phys. Rev. A} \textbf{84}, 043809 (2011).

\bibitem{Jona1993} F. Jona, G. Shirane, \emph{Ferroelectric crystals} (Dover, 1993).

\bibitem{DelRe1999} E. DelRe, M. Tamburrini, M. Segev, R. Della Pergola, A.J. Agranat, ``Spontaneous self-trapping of optical beams in metastable paraelectric crystals'' \emph{Phys. Rev. Lett.} \textbf{83}, 1954 (1999).

\bibitem{Kleemann1997} W. Kleemann, R. Lindner, ``Dynamic behavior of polar nanodomains in PbMg$_{1/3}$Nb$_{2/3}$O$_{3}$'', \emph{Ferroelectrics} \textbf{199}, 1 (1997).

\bibitem{Leuzzi2008} L. Leuzzi, T.M. Nieuwenhuizen, \emph{Thermodynamics of the glassy state} (Taylor \& Francis, 2008).

\bibitem{Mossa2004} S. Mossa, F. Sciortino, ``Crossover (or Kovacs) Effect in an Aging Molecular Liquid'', \emph{Phys Rev. Lett.} \textbf{92}, 045504-1 (2004).

\bibitem{Parravicini2012} J. Parravicini, A.J. Agranat, C. Conti, E. DelRe, ``Kovacs and inverse Kovacs effect in the optical scale-free regime'', submitted to \emph{Phys. Rev. Lett.} (2012).

\bibitem{Crosignani1998} B. Crosignani, E. DelRe, P. Di Porto, A. Degasperis, ``Self-focusing and self-trapping in unbiased centrosymmetric photorefractive media'', \emph{Opt. Lett.} \textbf{23}, 912 (1998).

\bibitem{Crosignani1999} B. Crosignani, A. Degasperis, E. DelRe, P. Di Porto, A. J. Agranat, ``Nonlinear optical diffraction effects and solitons due to anisotropic charge-diffusion-based self-interaction'', \emph{Phys. Rev. Lett.} \textbf{82}, 1664 (1999).

\bibitem{Ishai2004} P. Ben Ishai, C.E.M. De Olivera, Y. Ryabov, Y. Feldman, A.J. Agranat, ``Glass-forming liquid kinetics manifested in a KTN:Cu crystal'', \emph{Phys. Rev. B} \textbf{70}, 132104 (2004).

\bibitem{Gumennik2011} A. Gumennik, Y. Kurzweil-Segev, A.J. Agranat, ``Electrooptical effects in glass-forming liquids af dipolar nano-clusters embedded in a paraelectric environment'', \emph{Opt. Mater. Express} \textbf{1}, 332 (2011).

\bibitem{Note1} The final equilibration temperature is held through small-amplitude rapid adjustments that are omitted in the $S$ and $R$ curves.

\bibitem{DelRe2006} E. DelRe, A. Ciattoni, E. Palange, ``Role of charge saturation in photorefractive dynamics of micron-sized beams and departure from soliton behavior'', \emph{Phys. Rev. E} \textbf{73}, 017601 (2006).


\end{thebibliography}
\end{document}